\documentclass[sigconf,authorversion,nonacm]{acmart}
\usepackage{multirow}
\usepackage{bbm}
\usepackage{dsfont}
\usepackage{algorithm}
\usepackage{algorithmic}
\usepackage{float}

\newcommand{\revised}[1]{{#1}}
\newtheorem{example}{Example}

\usepackage{bbm}
\setcopyright{none}

\AtBeginDocument{%
  \providecommand\BibTeX{{%
    \normalfont B\kern-0.5em{\scshape i\kern-0.25em b}\kern-0.8em\TeX}}}

\begin{document}

\makeatletter
\def\@copyrightspace{\relax}
\makeatother

\title{Mitigate Position Bias with Coupled Ranking Bias on CTR Prediction}

\author{Yao Zhao}
\email{nanxiao.zy@antgroup.com}
\affiliation{%
  \institution{Ant Group}
  \streetaddress{569 Xixi Road}
  \city{Hangzhou}
  \country{China}
  \postcode{330100}
}

\author{Zhining Liu}
\email{eason.lzn@antgroup.com}
\affiliation{%
  \institution{Ant Group}
  \streetaddress{569 Xixi Road}
  \city{Hangzhou}
  \country{China}
  \postcode{330100}
}


\author{Tianchi Cai}
\email{tianchi.ctc@antgroup.com}
\affiliation{%
  \institution{Ant Group}
  \streetaddress{569 Xixi Road}
  \city{Hangzhou}
  \country{China}
  \postcode{330100}
}

\author{Haipeng Zhang}
\email{zhanghp@shanghaitech.edu.cn}
\affiliation{%
  \institution{ShanghaiTech University}
  \city{Shanghai}
  \country{China}
  \postcode{201210}
}

\author{Chenyi Zhuang}
\email{chenyi.zcy@antgroup.com}
\affiliation{%
  \institution{Ant Group}
  \streetaddress{569 Xixi Road}
  \city{Hangzhou}
  \country{China}
  \postcode{330100}
}

\author{Jinjie Gu}
\email{jinjie.gujj@antgroup.com}
\affiliation{%
  \institution{Ant Group}
  \streetaddress{569 Xixi Road}
  \city{Hangzhou}
  \country{China}
  \postcode{330100}
}

\begin{abstract}

Position bias, i.e., users' preference of an item is affected by its placing position, is well studied in the recommender system literature. However, most existing methods ignore the widely coupled ranking bias, which is also related to the placing position of the item. Using both synthetic and industrial datasets, we first show how this widely coexisted ranking bias deteriorates the performance of the existing position bias estimation methods. To mitigate the position bias with the presence of the ranking bias, we propose a novel position bias estimation method, namely gradient interpolation, which fuses two estimation methods using a fusing weight. We further propose an adaptive method to automatically determine the optimal fusing weight. Extensive experiments on both synthetic and industrial datasets demonstrate the superior performance of the proposed methods.
\end{abstract}



\maketitle

\section{Introduction}
Recommender systems (RS) play a key role in content recommendation~\cite{wu2021unbiased,saito2020unbiased,STAR}. It's the most effective way to alleviate information overloading for users and benefits content providers with more potential of making profits \cite{chen2020bias}. In many scenarios such as feeds and video recommendations, RS is often required to recommend multiple candidate items to the user at each query \cite{wu2021unbiased,zhao2019recommending}. These items are placed at different positions, and a user may scroll to the next page or click one (or more) items.  In the above multi-item recommendation scenarios, recent studies \cite{agarwal2019addressing,mehrabi2021survey} have found a critical factor that heavily affects users’ behaviors, i.e., position bias. Specifically, users may be more attracted by the items placed at the top positions rather than those at the bottom positions, regardless of the items’ actual relevance. Directly using these samples (i.e., implicit feedback data) to train a CTR estimation model will lead to inaccurate prediction of items’ true relevance. Because the existence of the position bias will amplify the CTR of the items placed at the top positions in the training set, hence deteriorating the recommendation performance.

Many works have been proposed to mitigate the position bias of CTR prediction in multi-item recommendation scenarios, such as position-as-module method \cite{guo2019pal}, knowledge distillation \cite{liu2020general}, adversarial learning \cite{wu2021debiasgan}, etc. In these methods, CTR is supposed to be fully modeled only by the position bias and the user’s true interest in the item. Specifically, the position bias can be modeled by user features and the position feature, and the user's true interest in the item can be modeled by user features and item features.

However, in multi-item recommendation scenarios, there is another source of bias beyond the position bias, which is neglected by the above works, the ranking bias. The ranking bias refers to the phenomena that the recommended items usually have a descending order of CTR \cite{mehrabi2021survey}, which universally exists because an RS tends to place candidate items with higher estimated CTR at top positions. Although both ranking bias and position bias are related to the position feature, they are caused by different reasons, i.e., ranking bias is caused by allocating user-interested items to the top positions, while position bias is caused by users' attention being more attracted by the top positions. Intuitively, the existence of ranking bias will affect the accurate estimation of position bias as it also lifts the average CTR of the top positions and depresses the average CTR of the bottom positions.

To investigate the coupled effect of ranking bias and position bias on CTR prediction, in this paper, we first perform analysis on both synthetic and industrial datasets and propose to study the position bias with coupled ranking bias problem. Specifically, we define the concept of position gradient, which reflects the CTR changes in response to the position changes, and we find the model misleads by the high CTR samples in the top positions and low CTR samples in the bottom positions, resulting in predicting a high and a low CTR score for an arbitrary item allocated at top positions and bottom positions, respectively. Therefore, the model will perform poorly when evaluating at a test set, as a test set for position bias evaluation is collected with random ranking strategy~\cite{joachims2017unbiased}, which would place an item at any position. We call it an overestimation of the position gradient problem. To overcome the overestimation, we propose a heuristic method, termed gradient interpolation, to depress the gradient of the position feature and to ease the overestimation significantly. Furthermore, we propose an adaptive solution to obtain an optimal interpolation coefficient efficiently if a few random ranking samples are available. Experiences on both synthetic and industrial datasets demonstrate the superior performance of the proposed method.

In summary, our main contributions are as follows:
\begin{itemize}
\item  We identify the widely existing problem of conventional position debias models, i.e., the coupled position bias and ranking bias would lead a conventional position-bias-aware model to overestimate the importance of the position feature, resulting in inaccurate user-to-item relevance score.

\item  We propose a novel position bias estimation method, namely gradient interpolation, to suppress the overestimation. It fuses a conventional overestimation model with an underestimation model to obtain a proper estimation. We also propose a method to determine an optimal fusing weight if a few random ranking samples are available.

\item  We perform experiments on two datasets and also two online recommendation scenarios, which demonstrate that the proposed method achieves consistent improvements.
\end{itemize}

\section{Related Work}
Targeting the problem of position bias, recent studies \cite{joachims2017accurately,yi2021debiasedrec, guo2019pal,chen2021autodebias,huang2021deep,jin2020deep,agarwal2019estimating,ai2018unbiased,vardasbi2020cascade,wu2021debiasgan,wang2018position,saito2020unbiased} estimate the effect of position in CTR prediction. A straightforward way is to inject the exposed position to network inputs~\cite{zhao2019recommending}, known as the position-as-feature (PSF) model. Usually a PSF model uses a default position in serving stage, to avoid determining the default position value, PAL~\cite{guo2019pal} is proposed to decompose CTR into position-dependent factor and position-independent factor, which assumes that position bias is a user-irrelevant multiplicative factor. However, later studies indicate that position bias also depends on user features~\cite{huang2021deep,yi2021debiasedrec}, which suggests the necessity to take user features into the modeling of position bias.  Another thread of research is based on inverse propensity weighting~\cite{rosenbaum1983central,ai2018unbiased,agarwal2019general,vardasbi2020cascade,joachims2017unbiased,zhu2020unbiased,zhang2020large}. It focuses on a general debiasing problem, while extreme propensities (i.e., the propensity scores close to 0 or 1) in position debiasing will cause the unstable learning problem~\cite{Fan2018Addressing}. Besides, knowledge distillation~\cite{liu2020general}, adversarial learning~\cite{wu2021debiasgan}, reinforce learning~\cite{huzhang2021aliexpress}, and evaluator-generator framework~\cite{bello2018seq2slate} are also introduced for position debiasing, however these methods are either in need of a great amount of samples, or difficult to convergent at real scenarios. 

\section{Overestimation of Position Gradient}

\begin{figure}[tb]
	\centering
	\includegraphics[width=3.6in]{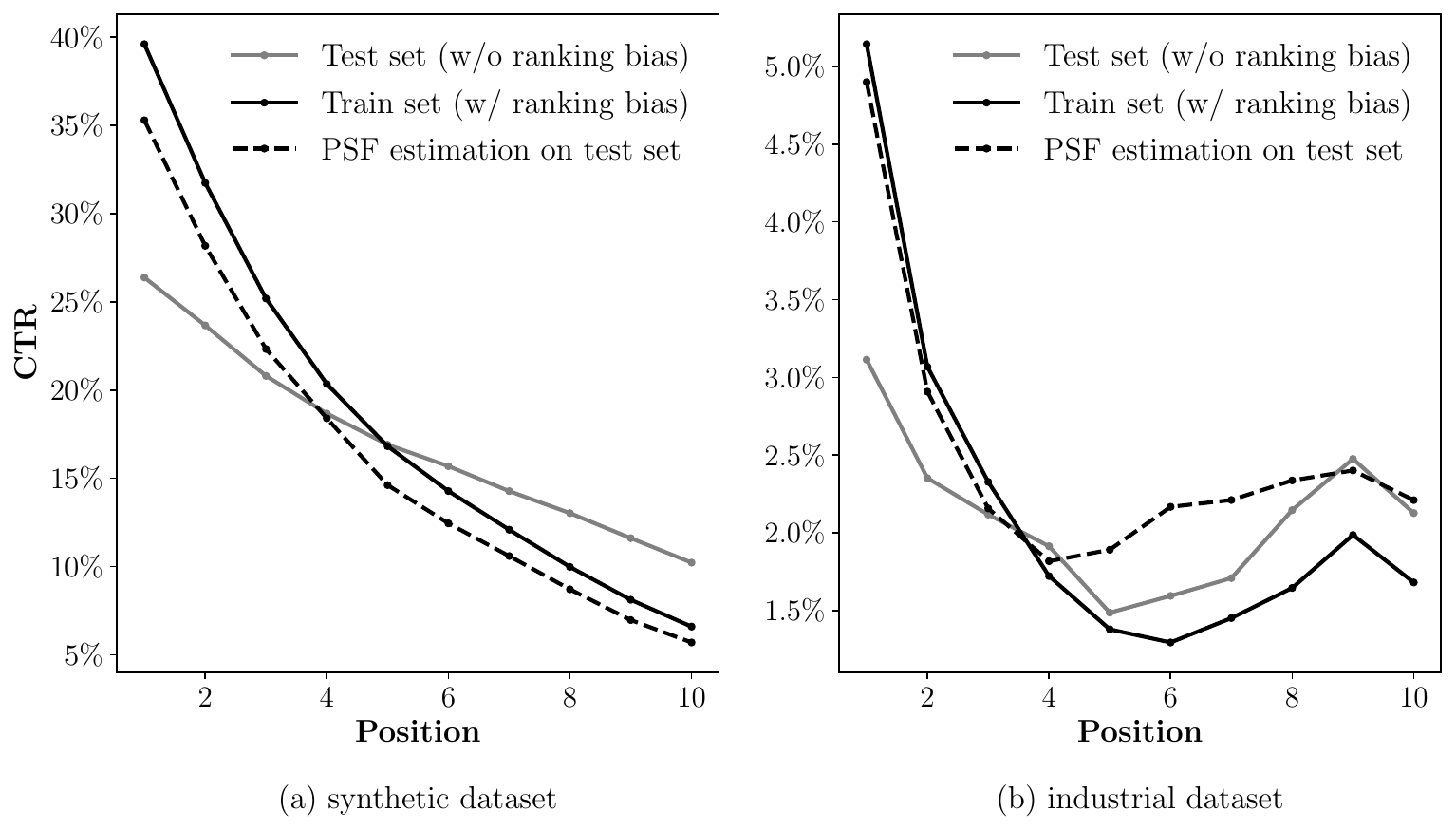}
	\caption{\revised{Average CTRs at different positions on various datasets are plotted. On both synthetic and industrial datasets (see Section \ref{dataset_explain} for details on datasets),  the training sets are collected using RS (with ranking bias), and the test set is collected under a fully random recommendation policy to eliminate the ranking bias. It is clear that with the coexistence of the ranking bias, the CTRs at the top positions are overestimated, and at the bottom positions are underestimated, which depicts the overestimation of the position gradient. The PSF method, fitted on the training set and evaluated on the test set, does not address the aforementioned problem.}}  
	\label{fig:bias_statistics}
\end{figure}

In this section, we discuss the limitation of the commonly used PSF methods in multi-item recommendation scenarios. We illustrate that the wide coexistence of the position bias and the ranking bias deteriorates the performance of the PSF methods with a synthetic example, and propose to study the \textbf{position gradient} overestimation problem.

\revised{In a multi-item recommendation task, the RS is trained using a previously collected dataset of user behaviors $\mathcal{D} = \{(u,i,k,y)\}$, where $u\in\mathcal{U}$, $i\in\mathcal{I}$ and $k\in\mathcal{K}$ are the features of the user, the recommended item and the position, respectively. $y$ is the label corresponding to each exposure. In the following, we assume $y\in\{0,1\}$ is the commonly used click label indicating whether the user clicked the recommended item. The position bias, for example, users are more likely to click the item placed on the top position, can then be formalized by the dependence between the position $k$ and label $y$ conditioned on the same user $u$ and the item $i$, i.e. $(y \not \! \perp \!\!\! \perp k | u, i)$. Similarly, the ranking bias, which reflects that in the collected dataset $\mathcal{D}$, user-preferred items are more likely to be placed in top positions can be expressed as $ (i \not \! \perp \!\!\! \perp k | u)$. We define the  position gradient to be the gradient of predicted CTR with respect to the position feature 
\begin{align*}
\mathrm{Position\ Gradient := }\ \mathbb{E}_{P \sim (i \perp \!\!\!  \perp k | u)}  \frac{\partial \left[y  \mid u, i, k\right]}{\partial k}  ,
\end{align*}
where the expectation is over a test set that is collected under a random recommendation strategy, to eliminate the ranking bias. Note that the position bias of different users is diverse, for example, a user visiting via a small screen device may result in more serious position bias. However, collecting such a test set without ranking bias requires randomly ranking the recommended items, whose results may not be preferable to the users and harm the user experience. Therefore in our work, we do not make the unrealistic assumption that the training set has no ranking bias. Instead, we assume that we can collect a small test set of data without ranking bias, and we analyze how the ranking bias in the training set affects the performance of the RS trained using the PSF methods using the test set. }

\revised{Figure \ref{fig:bias_statistics} plots the CTR against the position index for different datasets. }
For a well-fitted PSF model, its average estimated CTRs should approximate the CTRs of the test set on all positions. However, from the figure, we can find that there's significant discrepancy on test set position bias and estimated position bias via a PSF model. Specifically, compared with the CTRs of the test set, the estimated CTRs by PSF model are more sensitive to the positions, i.e., the gradient of the estimated CTRs of the PSF method trained on the training set, with respect to the position index, is steeper than that of the test set without ranking bias. As a result, changing the position feature from $k$ to $k+1$ would cause more CTR decay. This shows that the PSF model overestimates the position gradient.

We suppose that overestimation is mainly caused by learning with regularization, which rewards models of lower complexity \cite{2017Regularization,2016Understanding,2021Understanding}. Therefore, a model that attributes CTR to a single feature (e.g., the position feature) has lower complexity than the one that attributes it to a number of features (e.g., hundreds of item and user features).

To verify the above assumption, we start from a synthetic example using a linear model for analysis. We then qualitatively illustrate how the commonly coexistence of position bias and the ranking bias deteriorate the performance of the PSF methods with the following example.

\begin{example}
Assume for two exposures with features $v_a = [0.1, \dots, 0.1]$ and $v_b = [-0.1, \dots, -0.1]$. For the ease of presentation, we assume the ground truth click of each user is a Bernoulli distribution with expected determined by a linear model $pctr = \mathrm{sigmoid}\sum_{i=1}^{10}(w_i * v_i + w_0 * p)$, where $w_0=-1$ and $w_i=1$ for $\forall 1\le i\le 10$, and $p\in{0,1}$ for position 0 (the top one) and 1 (the bottom one). In the example, we use $loss^{emp} := (ctr - \hat{ctr})^2$ and $ loss^{reg} := 0.001 \sum_{i=0}^{10}w_i^2.$ to calculate the empirical loss and the regularization loss. The parameters of overestimation model are assumed to be $w_0=-1.1$ and $w_i=0.9$ for $\forall 1\le i\le 10$, which emphasizes the weight of the position feature.

\begin{table}[t]
	\caption{An synthetic example to illustrate overestimation of position gradient. An overestimation model could obtain even smaller loss than the ground-truth one within the biased samples. }
	\label{tbl.overestimation}
    \setlength\tabcolsep{5pt}
	\begin{tabular}{lccc}
		\toprule  
		Model &  $loss^{emp}$ &  $loss^{reg}$ &  $loss^{emp}+loss^{reg}$  \\
		\midrule 
	      Ground-truth model & \textbf{0} & 0.0110 & 0.0110  \\
            Overestimation model & 0.0004 & \textbf{0.0093} & \textbf{0.0097} \\
		\bottomrule 
	\end{tabular}
\end{table}

\end{example}

 As shown in Table \ref{tbl.overestimation}, overestimation of the weight of the position feature could obtain a lower loss with samples mixed with position bias and ranking biases. If we use a stochastic gradient descent method to optimize the parameters, the convergence model will be the model with smaller loss, i.e., the overestimation one.

To further study how L2 regularization affects the overestimation problem, we conduct the following experiment on our industrial dataset with different levels of L2 regularization. 

\begin{example}
We set L2 regularization coefficient to 1e-5, 1e-4 and 1e-3, respectively, and the overestimation ratio, to 1.391, 1.434 and 1.472, respectively, consistent with the expectation. Here, overestimation ratio is calculated by a ratio of $CTR_{
max}/CTR_{min}$ in a biased model to ground-truth $CTR_{max}/CTR_{min}$ in the random traffic, where $CTR_{max}$ and $CTR_{min}$ denotes the greatest and the least position-wise average CTR, respectively \footnote{We employ $CTR_{max}/CTR_{min}$ here rather than precise position gradient as it is easier to be calculated.}. The ratio measures how far the estimated position bias is away from the ground-truth position bias.

\end{example}

\section{Learning without Overestimation}

In the previous section, it was found that a conventional PSF model tends to overestimate the position gradient. To obtain an unbiased estimation, the position gradient in the PSF model should be depressed. However, it is challenging to implement with traditional training processes. An alternative approach is to create a position-unaware model by discarding the misleading position feature. This model will ignore the effect of the position feature, resulting in an underestimated position gradient of 0. Combining these two approaches can lead to an intermediate model that cancels out the overestimation and underestimation, as supported by Lagrange's mean value theorem.
We name the method as gradient interpolation, as it behaves similarly to the interpolation method in the field of numerical analysis.
Specifically, let the position gradient of the position-aware model be $g$, and the mixing weight for the position-aware model and the position-unaware model are $1-\epsilon$ and $\epsilon$, respectively, and the outputs of the position-aware model and the position-unaware model are $p_a$ and $p_u$, respectively. Then the output of the mixed model is $(1-\epsilon) p_a + \epsilon p_u$, and the position gradient is $(1-\epsilon)g$.

To determine an optimal $\epsilon$, we define an objective function that aims to find a $\epsilon$ that the expectation of the fused CTR score on each position is close to the expectation of ground-truth position bias (calculated from a small set of random ranking samples), i.e., minimizes the gap between the mediated CTR and the ground-truth CTR:

\begin{equation}
\begin{aligned}
\label{eq.min_rate}
argmin_{\epsilon} &  \Sigma( \epsilon p^a_i+(1-\epsilon)p^u_i-p^g_i)^2,\\
& s.t.   \quad 0<=\epsilon <=1 
\end{aligned}
\end{equation}
where $p^g_i$ is the average CTR of each position $i$, and $p^u_i$ and $p^a_i$ are average CTR of a position-unaware model and a PSF model in position $i$, respectively. 
Optimal solution for $\epsilon$ is:
\begin{equation}
\begin{aligned}
\label{eq.eps_solution}
\epsilon = \frac{\sum_i (p^g_i-p^u_i)(p^a_i-p^u_i)}{\sum_i (p^a_i-p^u_i)^2}. 
\end{aligned}
\end{equation}

To calculate the optimal weight, we should know $p_k$, $p^a_k$ and $p^u_k$ at first:
\begin{itemize}
	\item $p^g_k$: It is the ground-truth of the average CTR on $k$-\textit{th} position, therefore it can be directly calculated from the unbiased validation set.
	\item $p^a_k$: It is obtained by calculating the average predicted CTR of the unbiased validation set on $k$-\textit{th} position of $\mathcal{D}_u$ inferred by a PSF model $\mathcal{M}_a$, which is trained on the biased training set $\mathcal{D}_b$.
	\item $p^u_k$: Similar with $p^a_k$, $p^u_k$ can be obtained by the model $\mathcal{M}_u$ trained without position features. Fortunately, it can be approximated with the average of $p^a_k$ without training $\mathcal{M}_u$, i.e., $p^u_k=p^u=\frac{1}{K} \sum_{k=1}^{K} p^a_k$. This is because the output of $\mathcal{M}_u$ is irrelevant to the position feature, the outputs of different positions are approximately the same, and the mean predicted CTR of $\mathcal{M}_u$ and $\mathcal{M}_a$ is approximately the same as well.
\end{itemize}

\section{Experiments}

To evaluate the effectiveness of our proposed method, we conduct extensive experiments on both synthetic and industrial datasets. Empirical results show that our method gradient interpolation (denotes GI) outperforms the baseline methods with respect to the AUC metric and estimation error of position bias.

\subsection{Experimental Setup} \label{dataset_explain}

\noindent
\textit{\textbf{Datasets.}}
We use the following two datasets for evaluation, with dataset statistics shown in Table~\ref{tbl.dataset}.

\textit{\textbf{Synthetic Dataset}}: To the best of our knowledge, there is no public dataset that consists of samples from RS ranking traffic and random ranking traffic simultaneously. We first synthesize a dataset to conduct experiments, because a synthesized dataset is free of other biases (e.g., popularity bias, selection bias \cite{mehrabi2021survey} in a real-world RS.) and thus more ideal to model position bias and ranking bias. It contains 2 steps: (1) Generate random user and item features, and set position to $0$ (i.e., the first position), then score samples with a carefully designed function\footnote{$score=exp(sin(\Sigma_{i=1}^{32}x_i)+sin(\Sigma_{i=33}^{64}x_i)+cos(\Sigma_{i=1}^{32} x_{2i-1} - x_{2i})+ 0.01\Sigma_{i=1}^{32}sin(10x_i+10x_{i+32})-0.1p+0.1cos(\Sigma_{i=1}^{16}x_i-x_{i+16}+p)  + 0.1\sigma -3 )$, where $x_{i\in[1,32]}\sim uniform(0,1)$ is user features, $x_{i\in[33,64]}\sim uniform(0,1)$ is item features, $p\in [0,9]$ is position, $\sigma \sim uniform(-1,1)$ is noise, and $sin$ and $cos$ functions are applied for a non-monotone function. The score function is designed to depend on user-related features, item-related features, user-to-item features, a position feature and user-to-position features.}. 2) To generate samples to mimic RS ranking traffic, we rank items by scores generated from the previous step in descending order for each user to get final positions, then recalculate scores with the final positions using the above function, and relabel a sample as positive with the probability of the calculated score. For random traffic, we assign a random position to each sample.

\textit{\textbf{Industrial Dataset}}: We collect one-week traffic logs from an industrial RS -- it recommends nearby shops to users with feeds in our mobile app. The RS deployed a transformer-based deep model without the position feature. The training set is sampled from the previous 6 days, and the test set is sampled from the 7-\textit{th} day. In addition, 1\% of the traffic is collected under a fully random policy (i.e., using random scores to replace predicted CTRs of the baseline model) to make up a test dataset without ranking bias.

\noindent
\textit{\textbf{Comparison Methods.}}
We compare our method with the following baselines:

\textbf{BASE}: baseline model trained without a position feature.

\textbf{ST-PSF}~\cite{zhao2019recommending}: a PSF model with a shallow tower for the position feature.

\textbf{PAL}~\cite{guo2019pal}: a method that decomposes CTR estimation into position-independent factor and position-dependent factor.

\textbf{DPIN}~\cite{huang2021deep}: a method that combines all candidate items and positions for estimating CTR at each position.

\begin{table}
	\caption{Dataset statistics. }
	\label{tbl.dataset}
    \setlength\tabcolsep{3pt}
	\begin{tabular}{lcccc}
		\toprule  
		Dataset & \#samples& \#user & \#item  & \#click \\
		\midrule 
	    Synthetic training set & 8.4M & 839k & 100k & 1.5M \\
		Synthetic test set & 1.1M & 107k & 635k & 184k \\
		\hline 
		Industrial training set & 61.8M & 4.3M & 3.5M & 1.4M \\
		Industrial test set & 1.5M & 106k & 607k & 25k \\
		\bottomrule 
	\end{tabular}
\end{table}

\noindent
\textit{\textbf{Metrics.}} We use Area Under Curve (AUC) for the offline evaluation. We claim that position-related metrics (e.g., normalized discounted cumulative gain or mean reciprocal rank) are not proper for position bias evaluation, as position features are determinate, and we cannot reorder them arbitrarily. For the online evaluation, we use CTR as the metric, as it is closely related to user experience.

\noindent

\subsection{Implementation details}

We use transformer \cite{vaswani2017attention} as the backbone model architecture, and append a three-layer MLP with hidden size (1024, 512, 256) after transformer encoding to predict the CTR logit. The embedding size for the industrial dataset is 16. The shallow tower of ST-PSF is a one-layer MLP, the position module of PAL is similar with the baselines model but without transformer. The activation function is ReLU and the loss function is cross entropy. The optimizer is Adam with group lasso \cite{groupadam} with $\beta_1$=0.9, $\beta_2$=0.999, the learning rate is 0.001 and the batch size is 1,024. We use this optimizer to get a sparse embedding table in industrial RS. An early stopping strategy is applied to prevent over-fitting. 

In our gradient interpolation method, using two weighted models to estimate the CTR will consume double computation resources, in both training and serving stages. To improve the computation efficiency, we propose a novel randomization trick for accelerating. Instead of mixing the two models by weight, we use an equivalent sample fusion strategy. In detail, we randomly select a part of the samples in a mini-batch to assign random position features in the training stage, and then train and serve regularly. The ratio for randomization (denoted as randomization rate) is equal to $\epsilon$ in the mixed model. The sample fusion strategy is equal to the weighted mixing strategy because sampling a proportion of $\epsilon$ samples to assign a random position is equal to mixing a randomized-position model and a PSF model with weights $\epsilon$ and $1-\epsilon$.

\subsection{Offline Evaluation}\label{sec.offline}

Table~\ref{tbl.auc} shows the offline results of different methods, and we can observe that ST-PSF, PAL, and DPIN all fail to outperform BASE on the industrial test set, which is not surprising due to the overestimation of the position gradient. We use greedy searching to obtain the best weight, it can be observed that our model with the best weight is superior to the baseline models.  

 \begin{table}
	\caption{Offline evaluation of different methods. Bold indicates the best. The $\epsilon$ is obtained by greedy searching without using random ranking samples, to align with other methods.}
	\label{tbl.auc}
    \setlength\tabcolsep{3pt}
	\begin{tabular}{ccc}
		\toprule  
		Model & Synthetic Dataset & Industrial Dataset  \\
		\midrule
        BASE & 0.7240 & 0.6970 \\
        ST-PSF & 0.7318 & 0.6921 \\
        PAL & 0.7329 & 0.6965 \\
        DPIN & 0.7336 & 0.6943 \\
        \hline
        Ours & \textbf{0.7434} & \textbf{0.7077} \\
		\bottomrule 
	\end{tabular}
\end{table}

\begin{figure}[b]
	\centering
	\includegraphics[width=3.2in]{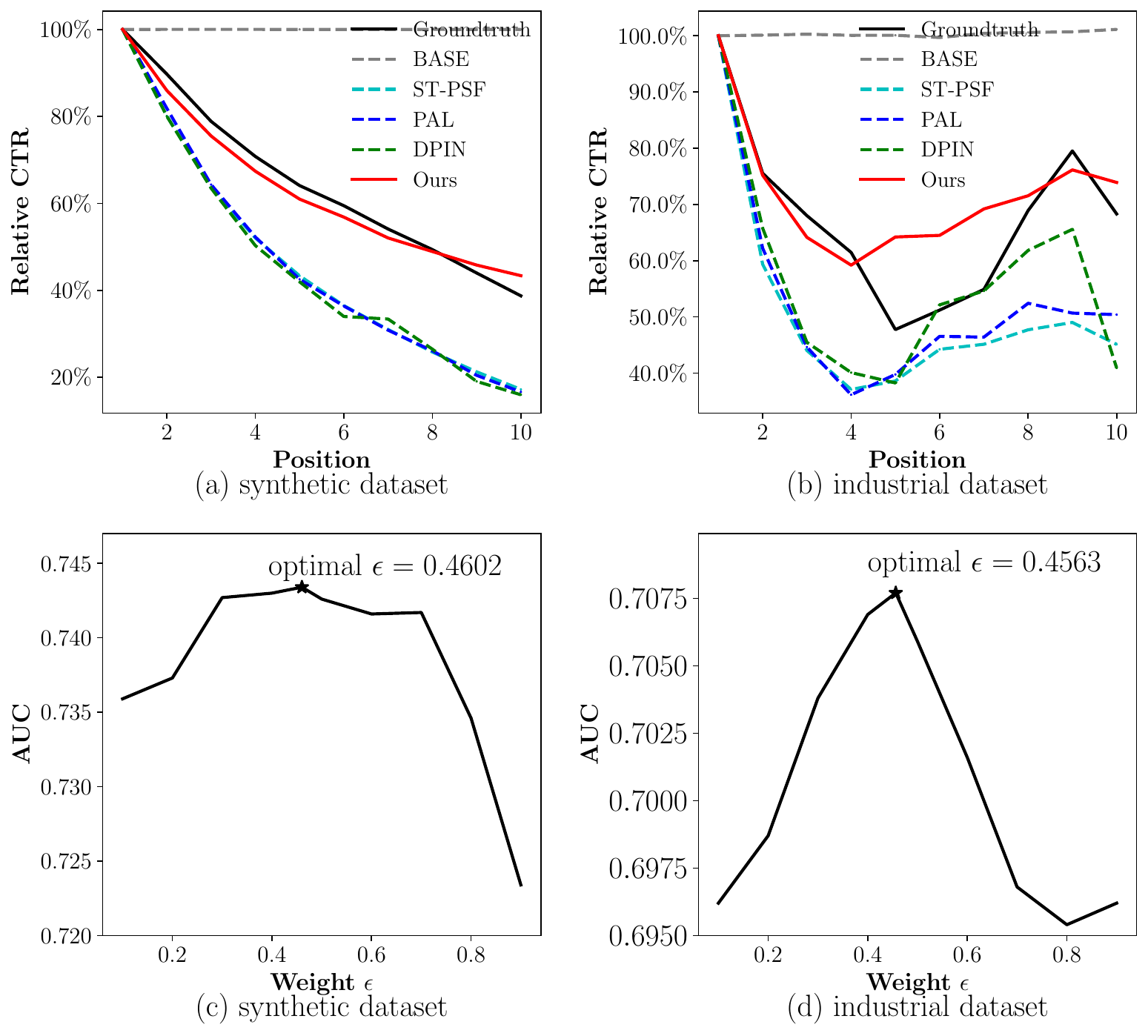}
	\caption{Evaluation of the error and AUC of different methods and different hyper-parameter $\epsilon$. In subfigures (a) and (b), the estimation error of our method is significantly less than other methods. Figures (c) and (d) prove that optimal $\epsilon$ can be obtained via equation \ref{eq.eps_solution} precisely without cumbersome greedy searching.}
	\label{fig.relative}
\end{figure}

To directly evaluate the CTR estimation errors, we plot the relative CTRs at different positions in figure \ref{fig.relative}. The relative CTR is calculated by dividing the average CTR of the first position, as we care more about the relative ratio instead of absolute values. The estimation error is the gap between model estimation and ground-truth estimation. It shows that our method has much less estimation error than other methods, especially in the synthetic dataset which perfectly matches the ground truth. The AUC of the optimal $\epsilon$ 
and searching grid $\epsilon$ is plotted in the sub-figure (c) and (d) in figure \ref{fig.relative}. It proves the effectiveness and efficiency of our solution to determine the optimal $\epsilon$.

\begin{figure}[h!]
	\centering
	\includegraphics[width=3.2in]{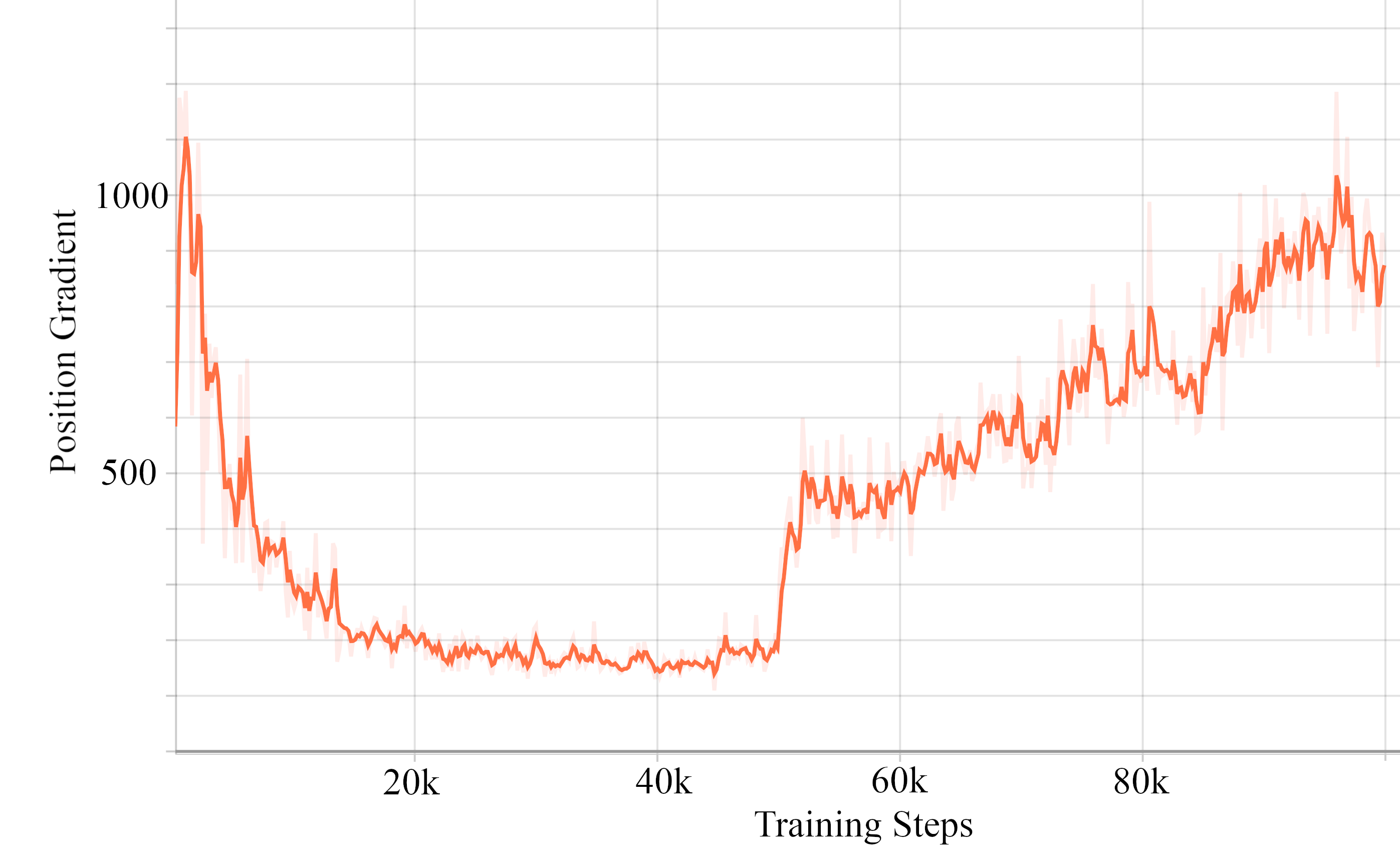}
	\caption{Position gradient on the industrial dataset. The position gradient of PSF (after the 50,000-\textit{th} step) is significantly larger than the one of the approximate ground-truth (before the 50,000-\textit{th} step).}
	\label{fig.grad}
\end{figure}

Furthermore, to quantify overestimation with position gradient directly, we train a model with the best weight obtained from the offline evaluation before the 50,000-\textit{th} step, the model can be approximately viewed as a ground-truth model. Then we change it to a PSF model by setting $\epsilon$ to 0. It shows in Figure ~\ref{fig.grad} that the position gradient is significantly increased after the change, which verifies the overestimation of the position gradient. 

\subsection{Online Evaluation}
To further verify the effectiveness of the proposed method, we launch two A/B tests on two online RS. For a fair comparison, we train all models on the same daily-updated dataset, and all methods share the same network architecture with the baseline model (a transformer-based~\cite{vaswani2017attention} backbone model). We do not deploy DIPN to online RS as it costs too many computational resources.  We use the optimal weight in the shop recommendation and use greedy searching for the goods recommendation. Table~\ref{table:online_ctr} shows that ST-PSF and PAL both perform worse than the base position-unaware model, while our proposed model shows significant improvement, which well demonstrates its effectiveness.

\begin{table}[h]
\caption{Online evaluation results. In two industrial recommendation tasks, our method achieves outperforms previous methods. Note that we don't test ST-PSF and PAL with the goods recommendation scenario as they are outperformed by the baseline method in the shop recommendation scenario.}
\setlength\tabcolsep{3pt}
\begin{tabular}{cccc|c}
\toprule
RS & BASE  & ST-PSF    & PAL    & Ours \\
\midrule
shop recommendation & 0.00\% & -2.99\% & -2.78\% & +3.43\% \\
goods recommendation & 0.00\% & - & - & +2.69\% \\
\bottomrule
\end{tabular}
\label{table:online_ctr}
\end{table}

\section{Conclusion}
In this paper, we consider the position bias problem in recommender systems. We first show that the coupled ranking bias leads to a position gradient overestimation problem. We then propose a novel position-bias-aware ranking method to address the overestimation of the position gradient problem. The proposed method, named gradient interpolation, alleviates the aforementioned overestimation problem. Offline and online experiments verify the effectiveness of our proposed method. For future work, we plan to determine the hyper-parameter weight without random ranking samples.

\newpage
\bibliographystyle{ACM-Reference-Format}
\bibliography{mpb}

\end{document}